# Preliminary study of $^{10}$Be/$^{7}$Be in rainwater from Xi'an by Accelerator Mass Spectrometry[*]

ZHANG Li (张丽)[1,2] and FU Yun-Chong (付云翀)[1,2†]

[1]State Key Laboratory of Loess and Quaternary Geology, Institute of Earth Environment, Chinese Academy of Sciences, Xi'an 710061, China.

[2]Xi'an AMS Center, Shaanxi Key Laboratory of Accelerator Mass Spectrometry Technology and Application, Xi'an 710061, China.

**Abstract:** The $^{10}$Be/$^{7}$Be ratio is a sensitive tracer for the study of atmospheric transport, particularly with regard to stratosphere-troposphere exchange. Measurements with high accuracy and efficiency are crucial to $^{7}$Be and $^{10}$Be tracer studies. This article describes sample preparation procedures and analytical benchmarks for $^{7}$Be and $^{10}$Be measurements at the Xi'an Accelerator Mass Spectrometry (Xi'an-AMS) laboratory for the study of rainwater samples. We describe a sample preparation procedure to fabricate beryllium oxide (BeO) AMS targets that includes co-precipitation, anion exchange column separation and purification. We then provide details for the AMS measurement of $^{7}$Be and $^{10}$Be following the sequence BeO$^{-}\rightarrow$Be$^{2+}\rightarrow$Be$^{4+}$ in the Xi'an- AMS. The $^{10}$Be/$^{7}$Be ratio of rainwater collected in Xi'an is shown to be about 1.3 at the time of rainfall. The virtue of the method described here is that both $^{7}$Be and $^{10}$Be are measured in the same sample, and is suitable for routine analysis of large numbers of rainwater samples by AMS.

## 1 Introduction

$^{10}$Be and $^{7}$Be are mainly produced in the stratosphere and upper troposphere by spallation reactions by secondary cosmic ray particles (principally neutrons and protons) with nitrogen and oxygen [1]. Shortly after production, these two isotopes become attached to aerosols and are transported through the atmosphere and ultimately, via dry/wet precipitation to the earth's surface [2]. The half-life of $^{7}$Be is 53.3 d, and it has been recognized as a natural tracer for atmospheric transport since the 1960s [3], and been widely used in the research of atmospheric dynamics in recent years [4-7]. $^{7}$Be deposited on the ground is also used as a tracer in studies of soil erosion [8]. $^{10}$Be is a long-lived isotope ($t_{1/2}$=1.387 Ma) [9, 10] and has a wide range of applications in areas such as atmospheric transport [11, 12], paleomagnetic field variations [13, 14] and as a monitor of solar activity [15].

In atmospheric transport applications, the usefulness of studying both $^{7}$Be and $^{10}$Be together stems from their common source as cosmogenic nuclides, their nearly identical geochemical behavior as isotopes of the same element, and the distinct contrast in their half-lives [1]. As an air mass moves towards the earth from the upper atmosphere (the region of highest production) the $^{10}$Be/$^{7}$Be ratio in aerosol particles invariably increases, as $^{7}$Be decays relatively quickly. In this case, the $^{10}$Be/$^{7}$Be ratio can be used as a chronometer and as a tracer in the study of stratosphere-troposphere exchange (STE) [16, 17]. Raisbeck and Yiou showed that the $^{10}$Be/$^{7}$Be ratio is a measure of residence time in the absence of air mass movement [16] (Fig. 1). The $^{10}$Be/$^{7}$Be in precipitation should bear information on where and when an air mass was formed. Thus one could trace the sources of air masses by measuring $^{10}$Be/$^{7}$Be ratios in rainwater [16]. $^{10}$Be and $^{7}$Be concentrations were determined in monthly rainfall collections at three sites across New Zealand (36° S to 45° S), from October 1996 to November 1998, by AMS and with γ-ray counting respectively, by Graham et al. [18]. They found that the average $^{10}Be/^{7}Be$ ratio of the rain collected from north New Zealand is 1.4-1.5 while that from south is 2.0. The seasonal variation of the $^{10}Be/^{7}Be$ ratio is higher in winter, which may reflect more active STE. Heikkila et al. focused on the $^{10}Be/^{7}Be$ ratios of rain samples collected from Dübendorf and Juunfraujoch, Switzerland, and probed the reasons for the differences [19].

---

[*] Supported by National Science Function of China (11205161) and CAS Key Technology Talent Program
[†] E-mail: fuyc@ieecas.cn

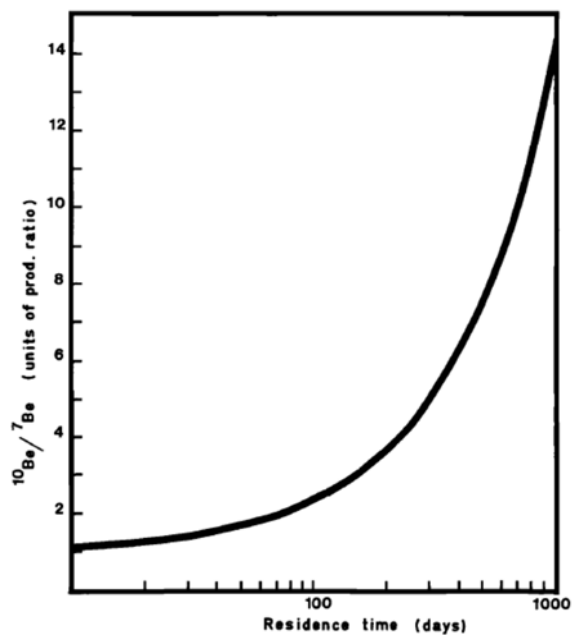

**Fig. 1. Rate of $^{10}Be/^{7}Be$ (in units of production ratio) as a function of residence time for an isolated air mass [16].**

Measurement with high accuracy and efficiency for $^{7}Be$ and $^{10}Be$ is crucial for the above studies. The measurement of $^{7}Be$ has been traditionally carried out by γ-ray counting, which is both economical and convenient. The measurement of $^{10}Be$ is almost exclusively performed by AMS. AMS also offers advantages for the measurement of $^{7}Be$, with inherently higher efficiency and sensitivity, as compared to γ-ray counting. AMS also has advantages for studies that utilize both $^{10}Be$ and $^{7}Be$. In this case it is especially convenient to measure both isotopes by the same instrument, which could eliminate inherent errors associated with the use of different equipments.

The Xi'an AMS Center has focused on tracing geomagnetic field changes with $^{10}Be$ in loess [14][20, 21], reconstructing paleoprecipitation [22], and exposure/burial chronology [23]. These studies require high precision $^{10}Be$ measurements from the Xi'an AMS instrument. Prior to 2007, our studies involving $^{7}Be$ and $^{10}Be$ in rainwaters for the reconstruction of paleoprecipitation, relied on combining the Xi'an AMS $^{10}Be$ data with $^{7}Be$ results reported by others. More recently we have performed our own $^{7}Be$ measurements with the γ-ray counting method, in collaboration with the Technical University of Denmark. In this case the samples were first analyzed for $^{7}Be$, then reprocessed to BeO targets for $^{10}Be$ AMS (personal communication, Dr. Kong Xianghui). The motivation for developing a single procedure to measure both $^{7}Be$ and $^{10}Be$ by AMS is clear, and this article discusses our progress in developing such a methodology. The method is expected to be especially useful for studies of beryllium isotopes in the atmosphere.

## 2 Rainwater sample preparation for $^{7}Be$ and $^{10}Be$ measurement at Xi'an AMS

Rainwater samples were collected from the roof of the Xi'an AMS Center building (34.2237°N,

109.9000°E), Institute of Earth Environment, Chinese Academy of Sciences (IEECAS), located in Xi'an, Shaanxi, P.R.China.

For rainwater samples, a typical sample preparation procedure for $^7Be$ analysis for γ-ray counting includes filtration, the addition of a $FeCl_3$ solution, and the addition of a $NH_3·H_2O$ solution to precipitate $Fe(OH)_2$ [24-26]. The γ-ray counting method does not require the removal of B, which can cause significant interferences in AMS beryllium-10 measurements. Therefore we developed and tested a sample pretreatment scheme for boron removal. This included the use of an ion exchange resin column to extract and purify Be, and remove any impurities that might include B [18][27]. An earlier procedure designed to process rainwater samples for $^{10}Be$ analyses failed to achieve an optimal recovery rate, due to repeated centrifugation steps. Hence we refined the process, and derived the following procedure, shown diagrammatically in Fig. 2.

(1) The rainwater sample was weighed (typical weight of ~11 kg). A known quantity of $^9Be$ was added to the sample at the time of collection. The sample was filtered to remove solid particles. $HNO_3$ was added to the solution to adjust to pH=2.

(2) About 2 ml $FeCl_3$ (10 mg/ml) solution was added to the sample. After 1 hour allowing for equilibration, $Fe(OH)_3$ was precipitated by adding $NH_3·H_2O$, allowing time (~20 min) for the reaction.

(3) The large amount of the supernatant was siphoned off with a rubber suction bulb and latex tube. The remaining solution was centrifuged, and the supernatant was poured off. $HNO_3$ was added to dissolve the precipitant. 2 mol/L NaOH was added to adjust the pH to 14. The solution was centrifuged and the supernatant was transferred to the centrifuge tube. $HNO_3$ was added to the supernatant to adjust the pH to 8-9. The solution was centrifuged and the supernatant was poured off. The precipitant was then recovered.

(4) Beryllium was separated and extracted by anion exchange chromatography, and leached with 9 mol/L HCl (about 80 ml).

(5) Ammonium hydroxide was added to the $^{10}Be$ fraction to precipitate beryllium hydroxide gel at pH=8-9, and the hydroxide was oxidized by ignition at 900°C in quartz crucibles.

(6) The beryllium oxide was mixed with niobium powder and pressed into target holders prior to measurement on the AMS.

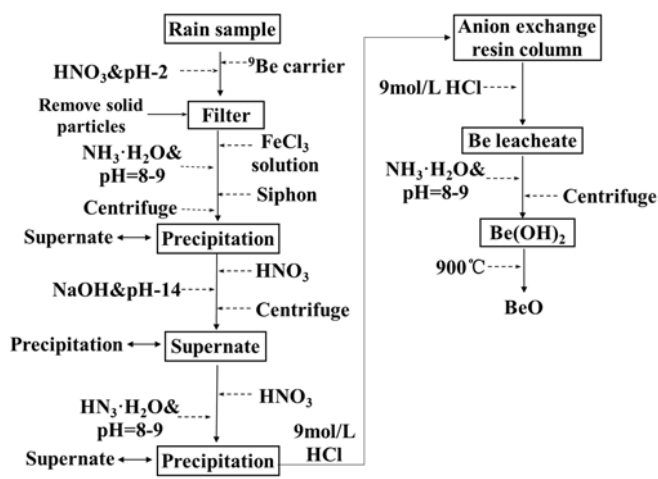

Fig. 2. Flow chart of rain sample preparation

The standard solutions and reagents used in the sample preparation were as follows: 1) a $^9Be$ standard solution from the NCS Testing Technology Co. Ltd (NCS) with a $^9Be$ concentration of 1000 mg/L; 2) a $^{10}Be$ standard (SRM 4325) from the National Institute of Standards and Technology (NIST), with a $^{10}$Be/$^9$Be ratio = 2.68×10$^{-11}$; 3) purified H$_2$O, prepared with a Milli-Q ultrapure water system (resistivity = 18.2 MΩ·cm, 25°C); 4) reagent grade HNO$_3$, HCl, NH$_3$·H$_2$O, NaOH and FeCl$_3$. The anion exchange resin used was Dowex1W-X8 (Cl$^-$), produced by Merck Co. Bio-rad 1.0 cm×20 cm ion exchange columns were used.

## 3 Establishment of method for AMS measurement and discussion

The preliminary measurement of $^7$Be-AMS was performed on the 3 MV multi-element AMS instrument at the Xi'an AMS Center, IEECAS. The ion injection energy used was 35 keV and a fast, 100 Hz cycling frequency was routinely employed *(Fast sequential injection FSI)*. The terminal voltage for the $^{10}$Be measurement is 2.5 MV, and Ar is used as the stripping gas. The high-energy system is equipped with a 115° analysis magnet, a 500 nm Si$_3$N$_4$ film for $^{10}$Be measurement, a 65° electrostatic analyzer and an additional 30° analysis magnet. The dual-anode gas ionizer detector was filled with 26 mbar isobutane and equipped with a 1500 nm PET window. The first anode plate (ΔE) is 6.0 cm in length and the second plate ($E_f$) is 9.5 cm. Other details of the system are described elsewhere [28-30].

Recently, a new method for $^{10}$Be-AMS analysis was tested at the Xi'an-AMS Center by Fu et al. that employs BeF$_3^-$ super-halogen anions [31]. These anions inherently suppress $^{10}$B interferences by nearly 5 orders of magnitude because the BF$_3^-$ anions rarely form. The resulting $^{10}$B suppression factor is still not as high as that achieved with energy degrader foils, but the suppression of $^{10}$B and separation from $^{10}$Be separation in the final ionization detector was found to be sufficient for $^{10}$Be$^{2+}$ to be counted directly using a small AMS. Because the degrader foil was not used, the high energy $^{10}$Be transmission efficiency was significantly improved, from ~30% to ~100% in the Xi'an-AMS. The advantage of using super-halogen anions also applies to $^7$Be measurement, for the yield of LiF$_3^-$ is negligible, in accordance with the electron configuration of Li, which leads to the preferential formation of LiF$_2^-$. Hence, potential interference from $^7$Li is eliminated during sample pretreatment, allowing for the AMS measurement of $^7$Be. It will be very convenient for both $^7$Be and $^{10}$Be to be determined by AMS using this analytical method, from the same target and at the same time. In other respects, $^7$Be behaved as $^9$Be and $^{10}$Be in the high energy analysis segment of the AMS, after the 115° analysis magnet. Hence, no commercial standard samples are required to correct $^7$Be by $^{10}$Be, often referred to as an interior correction. A significant limitation of this method at this stage however, are the chemical properties of fluoride ion conductors, which are very reactive and lead to enhanced memory effects in the ion source. Thus, further improvements are required to make the method viable for routine measurements.

Hence, we chose BeO as the target material for this study. The first AMS measurement of $^7$Be was made with a small AMS (≤3 MV) by Raisbeck and Yiou [32]. They also used a carbon foil, as in the case of $^{10}$Be. Although both methods employed a carbon foil to separate isobars, the principles were different. In the case of $^{10}$Be and $^{10}$B, the intent of the foil was to exploit the different energy losses experienced by the two isobars as they passed through the foil. In the case of $^7$Be and $^7$Li, the foil served to strip Be$^{2+}$ ions to Be$^{4+}$ ions. This eliminated $^7$Li ions since the maximum valence of a Li ion is +3. This strategy resulted in a very low blank value ($^7$Be/$^9$Be＜10$^{-15}$). We focused on this feature of the analysis for the Xi'an-AMS, to utilize Be$^{4+}$ ions to eliminate Li interference.

There are two ways to obtain Be$^{4+}$. One is to directly strip the ions to +4 using the argon stripper in the terminal of the accelerator: BeO$^-$→Be$^{4+}$; the other is to do the second stripping using the foil after the ions pass through the analysis magnet at the high energy system: BeO$^-$→Be$^{2+}$→Be$^{4+}$. Our first step was to use $^9$Be ions to perform a feasibility study, to compare these two ways of producing Be$^{4+}$ ions.

For the first experiment, we stripped Be to +4 ions directly at the terminal under the following conditions. The terminal voltage was adjusted to 2.5 MV and the argon pressure in the stripper was set to about $8.3\times10^{-3}$ mbar. These settings were chosen from routine measurements. The stripping efficiency for different valences was obtained with $^9$Be ions. At 2.5 MV, BeO$^-$→Be$^{2+}$ is about 47%, BeO$^-$→Be$^{3+}$ is about 2%, and BeO$^-$→Be$^{4+}$ is about 0.033% (almost zero). Therefore, we ruled out this scheme.

For the second experiment we stripped Be$^{2+}$ ions in the high energy analysis segment of the AMS using the existing 500 nm Si$_3$N$_4$ foil. The energy of the Be$^{2+}$ ions are about 6 MeV. The stripping efficiency of Be$^{2+}$→Be$^{3+}$ is about 31% through the Si$_3$N$_4$ foil. The Be$^{2+}$→Be$^{4+}$ stripping efficiency is about 3%, nearly one order of magnitude lower than Be$^{2+}$→Be$^{3+}$. For the present preliminary study we did not consider detecting $^7$Be using the $^7$BeO$^-$→$^7$Be$^{2+}$→$^7$Be$^{3+}$ sequence concerning the potentially excessive flux of $^7$Li$^{3+}$ ions entering the detector, because we had not considered special steps to remove Li during sample processing. This will be considered in future to obtain significantly higher measurement efficiency, after our data acquisition system is upgraded to capture complete $\Delta E/E_f$ 2D spectra. At this level, the total efficiency can still be analyzed by AMS so we chose it to test this method further.

We accelerated BeO$^-$ ions from the ion source, and selected +2 charge state coming out of the accelerator. We performed the secondary stripping to +4 using the Si$_3$N$_4$ foil for both $^7$Be and $^{10}$Be. This gave us $^{10}$Be/$^7$Be results from a single sample using one method. Since we intend to make a batch correction of $^7$Be with $^{10}$Be, we maintain consistency for the measurement of both isotopes. This means that we measure $^{10}$Be differently from past routine measurements which use Be$^{3+}$ ions. While we sacrifice some beam efficiency by selecting the +4 charge state, the change has no effect on our ability to distinguish between $^{10}$Be and $^{10}$B.

Our first test was to make measurements of the $^{10}$Be standard sample (NIST SRM4325 $^{10}$Be/$^9$Be =$2.68\times10^{-11}$) and $^{10}$Be in a rainwater sample. The $^{10}$Be/$^9$Be ratios for standard Be$^{2+}$→Be$^{3+}$ measurements and Be$^{2+}$→Be$^{4+}$ measurements are $8.35\times10^{-12}$ and $7.98\times10^{-13}$ respectively for the standard; while the $^{10}$Be/$^9$Be ratios are $2.61\times10^{-13}$ and $2.54\times10^{-14}$ for the rainwater sample.

Next we made $^7$Be measurements on the rainwater sample. $^9$Be$^{2+}$ is widely separated from $^7$Be$^{2+}$ after passing through the high energy analysis magnet. The offset Faraday Cup for our system is set for $^{26}$Al measurements so that we are currently unable to measure $^7$Be and $^9$Be in FSI mode, whereas we can make FSI measurements for $^9$Be and $^{10}$Be. We take the mean value of $^9$Be current before and after the $^7$Be measurement in the detector to normalize the $^7$Be counts. $^7$Be is injected by DC mode into the accelerator, passing through the Si$_3$N$_4$ film and then the 65° ESA and an additional 30° magnet into the detector. With terminal voltage kept at 2.5 MV, all high energy system analyzers had to be adjusted for selecting $^7$Be and $^{10}$Be detection. The spectrum of $^7$Be$^{4+}$ ions was simulated by the Srim program in the absence of any suitable pilot beam.

The $^9$BeO$^-$ beam current of the rainwater sample agreed well with that of the $^{10}$Be standard by using the new chemical preparation procedure; they were all close to 1 μA.

The $\Delta E$ energy spectrum of $^7$Be, in coincidence with correct signals detected in the $E_f$ anode of the gas detector, is shown in Fig. 3a. The $\Delta E$ spectrum for $^{10}$Be is shown in Fig. 3b. We observed a small shift between the two peaks. The peak position of $^7$Be is at about at channel number 262 while that $^{10}$Be is at about channel 283. As the terminal voltage is 2.5 MV, the total energy of $^7$Be $E_{7\text{-Total}}$ is estimated at 5.773 MeV, while the total energy of $^{10}$Be $E_{10\text{-Total}}$ is estimated at 5.976 MeV. The energy loss for each isotope through the Si$_3$N$_4$ film, window and the $\Delta E$-$E_f$ anodes were simulated using the Srim program. The energy losses at the $\Delta E$ anode were $\Delta E_7 \approx 3.487$ MeV and $\Delta E_{10} \approx 3.697$ MeV; the residual energies at the Ef anode were $E_{f\,7} \approx 1.775$ MeV and $E_{f\,10} \approx 1.719$ MeV. The measurement and simulation agree to each other very well.

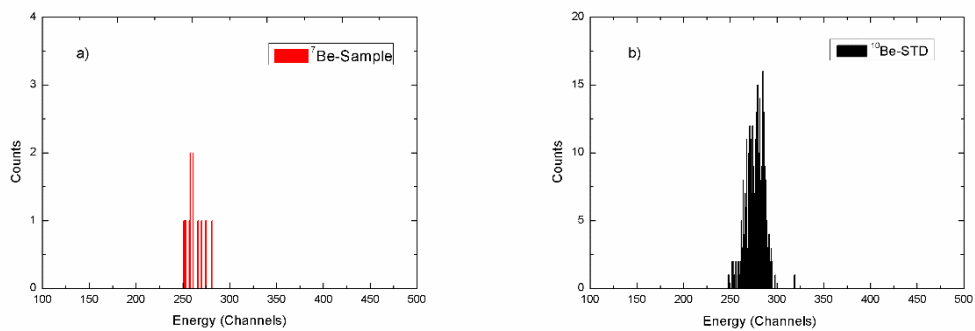

**Fig. 3. The ΔE spectrum coincidence with $E_f$ using $Be^{4+}$: a) $^7Be$ spectrum; b) $^{10}Be$ spectrum.**

We have thus established $^7Be$-$^{10}Be$ sample preparation and measurement methods for rainwater samples, using a single sample by AMS. This work also marks the first $^7Be$ measurements at the Xi'an-AMS laboratory. For the rainwater sample in this study, we carried out the normalization by current and decay correction for $^7Be$. Total counts within 600 s were 13 ions and the average current of $^9Be$ was about 0.71 μA. The result was about 18.31 atoms/μA within 600 s after normalization. Given that the rainwater sample was collected on Aug 12, 2015 and we did the measurement on Oct 3, an interval of about 53 days; the corrected $^7Be$ value is 36.62 ions at the time of sample collection. In addition, total counts of $^{10}Be$ were 39 atoms within 600 s and the average current of $^9Be$ was about 0.82 μA. The result is about 47.56 atoms/μA within 600 s after normalization processing. For the same chemical procedural blank, there are no counts for both $^{10}Be$ and $^7Be$ within 600 s. Then the $^{10}Be/^7Be$ ratio in the rain sample from Xi'an on Aug 12, 2015 was about 1.3. For comparison we find a common range of published $^{10}Be/^7Be$ rainwater values between 1.3-2.2 [18, 25, 33-35], with some higher values [19]. All of these studies used a combination of γ-ray counting for $^7Be$ and AMS for $^{10}Be$.

## 4 Conclusions and prospects

We report here a new AMS method for the measurement of $^7Be$ and $^{10}Be$. This method involves both a refined sample preparation method and AMS analysis method, using beryllium ions in the +4 charge state. We present results from a rainwater sample collected in Xi'an collected on August 12, 2015.

The use of $BeO^-$ as the extracted beam from the ion source allows us to perform $^{10}Be$ corrections with available standards. However, we have yet to find a suitable commercial $^7Be$ standard sample, and the correction method for $^7Be$ needs to be further optimized. In addition, further automation upgrade to the AMS instrument would significantly improve the new method. The $^7Be$ and $^{10}Be$ measurements can be performed automatically under slow injection conditions *(Slow sequential injection SSI）*.

There are a great number of sites in China that could be studied with this technique. The geographic and orographic setting of China is unique, from sites at sea level to the east, to the high plateau in the west, making China an especially valuable testing ground to understand air mass movements through space and time. This method is especially well-suited to establish $^{10}Be/^7Be$ time series, to capture anything from diurnal, to seasonal, or even annual variations.

**Acknowledgement**

We gratefully acknowledge the assistance provided by Zhao Guo-Qing in the sample collection. We also thank Dr. Zhao Xiao-Lei, Dr. George Burr, Prof. Hou Xiao-Lin, Prof. Wu Zhen-Kun and Dr. Dong Gou-Cheng for valuable discussion in this work.